%% file: epl_final.tex
\documentstyle[epsfig]{europhys}
\input euromacr

\begin{document}
%
%
%
\euro{}{}{}{}
\Date{}
\shorttitle{T.CALARCO \etal ARE VIOLATIONS TO TBI THERE WHEN SOMEBODY LOOKS?}
%
%
%
\title{Are violations to temporal Bell inequalities there\\
when somebody looks?}
\author{T. Calarco\inst{1},
     M.Cini\inst{2}, \And R. Onofrio\inst{3}}
\institute{
     \inst{1} Institut f\"ur Theoretische Physik - Universit\"at Innsbruck,\\
	Technikerstra\ss e 25/2, Innsbruck, Austria 6020\\
     and ECT*, European Centre for Theoretical Studies in Nuclear Physics\\
        and Related Areas, Strada delle Tabarelle 286, Villazzano (Trento), 
	Italy 38050\\
     \inst{2} Dipartimento di Fisica - Universit\`a di Roma ``La Sapienza''\\
        and INFN, Sezione di Roma, P.le Aldo Moro, 2, Roma, Italy 00185\\
     \inst{3} Dipartimento di Fisica ``G. Galilei'' - Universit\`a di Padova\\
        Via Marzolo 8, Padova, Italy 35131\\
  	and INFM, Sezione di Roma 1, P.le Aldo Moro, 2, Roma, Italy 00185} 
%
%
\rec{}{}
%
%
%
\pacs{
\Pacs{03}{65Bz}{Foundations, theory of measurement, miscellaneous theories
(including Aha-ronov--Bohm effect, Bell inequalities, Berry's phase)}
\Pacs{42}{50Lc}{Quantum fluctuations, quantum noise, and quantum jumps}
\Pacs{85}{25Dq}{Superconducting quantum interference devices (SQUIDs)}
      }
\maketitle
%
%
%
\begin{abstract}
The possibility of observing violations of temporal Bell inequalities,
originally proposed by Leggett and Garg as a mean of testing the quantum
mechanical delocalization of suitably chosen macroscopic bodies, is
discussed by taking into account the effect of the measurement process. 
A general criterion
quantifying this possibility is defined and shown not to be fulfilled by
the various experimental configurations proposed so far to test
inequalities of different forms.
\end{abstract}
%
%
%
%
%

\section{Introduction}
The validity of quantum mechanics at the macroscopic level is still an open 
question.
Leggett and Garg \cite{LEGGETT} have challenged this question by 
proposing 
laboratory tests 
aimed at comparing, in a macroscopic domain, the predictions of  a set of 
theories incorporating realism and non-invasivity, two properties not 
shared by quantum mechanics, versus quantum mechanics itself.
The test involved certain inequalities, called {\it temporal} Bell 
inequalities in analogy to the well-known {\it spatial} ones \cite{BELL}. 
These last have been already experimentally studied \cite{ASPECT} 
to understand the validity of  quantum mechanics against {\em local 
realism} at the {\em microscopic} level. 
Temporal Bell inequalities instead, involving correlation probabilities 
between subsequent measurements of the same observable, hold in 
{\em realistic }
theories but are violated under certain conditions by quantum mechanics.
The general ingredients required to discuss temporal Bell inequalities, 
regardless of the concrete scheme used, are three: 
{\it (a)} the possibility to have a coherent, quantum evolution of the state; 
{\it (b)} a dichotomic observable, already discrete or derived 
by a continuous one;
{\it (c)} different-time correlation probabilities between subsequent 
measurements of the dichotomic observable.
Leggett and Garg originally proposed use of an rf-SQUID, 
further analyzed in \cite{TESCHE}, while more recently other 
configurations have been discussed, such as a two-level atomic 
system undergoing optically driven Rabi oscillations \cite{PAZ}, and Rydberg 
atoms interacting with a single quantized mode of a superconducting 
resonant cavity \cite{SANTOS}. 
The proposal of Leggett and Garg has been criticized 
\cite{CRITICISM} focusing mainly on the assumptions made in the 
macrorealistic approaches.    

In this Letter we instead consider the full quantum mechanical 
predictions for the various proposed experiments. 
The main point of our analysis is that, as in any quantum mechanical 
prediction, probability distributions for the outcome of subsequent 
measurements have to be evaluated and compared to the analogous quantities 
actually measurable in laboratory. 
In practice, we will restrict ourselves to the evaluation 
of average and variance of the observed quantities.
While the average value is {\it necessary} to define the violation itself, its
variance is crucial to establish if the corresponding experimental resolution 
is {\it sufficient} 
to assign in an unambiguous way the value of the dichotomic observable 
\cite{ONCAL}. We will check if, under the experimental conditions which would 
lead to a contradiction between quantum and classical predictions 
for the correlation probabilities, the actual resolution can be sufficient to
distinguish the two values of the dichotomic observable --
otherwise it would be impossible to state that the quantum mechanical
probabilities refer to well--defined observables.
This feature was not taken into account in 
any of the previous papers of other authors on the subject \cite{CRITICISM}, 
and constitutes the main point of our
work. Indeed, we have already addressed this point elsewhere \cite{ONCAL,KONST}
referring to specific physical schemes; the novelty of the present contribution
is represented by the generality of our consideration, which apply to all
experimental proposals put forward so far.

\section{Quantum measurements on bistable systems}
We begin by briefly reviewing the conventional analysis of quantum
measurements on a generic bistable system undergoing oscillations with period
$\tau$ between the two states $|+\rangle$ and $|-\rangle$, according to the 
time evolution operator $\hat{U}(t_{\rm b},t_{\rm a})$. 
The effect of a measurement of the dichotomic observable 
$\hat{X}$  is, as usual, represented through a non-unitary measuring operator 
\begin{equation}
\label{FILTERDISCR}
\hat{\Pi}_{X}^\dagger=\hat{\Pi}_{X}\;,\quad\hat{\Pi}_{X}^2=1\!\mbox{l}\;,\quad
\hat{\Pi}_{X}|x\rangle=\delta_{Xx}|x\rangle\;,
\end{equation}
where $|x\rangle=|\pm\rangle$ and $X$ is one of the two eigenvalues 
$\pm |X|$. 
Given some initial preparation $|X_{\rm a} \rangle$ at time $t_{\rm a}$, 
after measurements performed at times $t_{\rm b}$, $t_{\rm c}$ with results
$X_{\rm b}$, 
$X_{\rm c}$ respectively we get  
\begin{equation}
\label{SIGMATBPLUS}
|x(t_{\rm b}^+)\rangle_{X_{\rm a}X_{\rm b}}=\hat{\Pi}_{X_{\rm b}} 
\hat{U}(t_{\rm b},t_{\rm a})|X_{\rm a} \rangle\;; \qquad
|x(t_{\rm c}^+)\rangle_{X_{\rm a}X_{\rm b}X_{\rm c}}=\hat{\Pi}_{X_{\rm c}}
\hat{U}(t_{\rm c},t_{\rm b})|x(t_{\rm b}^+)\rangle_{X_{\rm a}X_{\rm b}}\;.
\end{equation}
The corresponding correlation probabilities are 
\begin{equation}
\label{PABPBC}
p^{\rm ab}_{X_{\rm a}X_{\rm b}}=
||x(t_{\rm b}^+)||^2_{X_{\rm a}X_{\rm b}}\;;\qquad
p^{\rm bc}_{X_{\rm a}X_{\rm b}X_{\rm c}}=
||x(t_{\rm c}^+)||^2_{X_{\rm a}X_{\rm b}X_{\rm c}}\;,
\end{equation}
and their variances, whose square roots are called effective uncertainties, 
are 
\begin{equation}
\label{DSIGMAABBC}
\left(\Delta X^{\rm ab}_{\rm eff}\right)^2={\sum_{x} \left( X_{\rm b}-x
\right)^2 p^{\rm ab}_{X_{\rm a}x}}\;;\qquad
\left(\Delta X^{\rm bc}_{\rm eff}\right)^2={\sum_{x} 
\left( X_{\rm c}-x\right)^2 p^{\rm bc}_{X_{\rm a}X_{\rm b}x}}\;.
\end{equation}
Their value is dictated by the laws of quantum mechanics, regardless of
the size of the ensemble over which measurements are
performed. From the definition Eq.~(\ref{DSIGMAABBC}) 
it follows that $0\leq\Delta X_{\rm eff}\leq 2|X|$ 
at any times. On the other hand, a resolution large enough is required 
in order to assign to the dichotomic observable a non-zero value without
ambiguity. In other words, one has to take as non-ambiguous only 
 measurements for which a condition 
$\Delta X_{\rm eff}\leq\xi$ is satisfied, the value of 
the threshold effective uncertainty $\xi$ defining the 
resolution criterion. A null value of $\xi$ means that only
infinite-resolution measurements are taken as non-ambiguous; when 
$\xi =2|X|$ no discrimination is made between measurements with 
or without a resolution power sufficient to distinguish the two states 
$|+\rangle$ and $|-\rangle$. 
Any of the criteria commonly used in resolution studies \cite{HANDBOOK} 
can be related to a value of the threshold 
between these two extremes. For example, the well-known half-width 
criterion corresponds to $\xi=\xi_{1/2}\equiv (2\ln 2)^{-1/2}|X|$: 
two probabilities distribution centered at 
$\pm|X|$ can be resolved if and only if
their variance satisfies $\Delta X<\xi_{1/2}$. A less stringent criterion 
would be for instance to require $\Delta X<|X|$.

\section{Temporal Bell inequalities}
Temporal Bell inequalities are formally derived by assuming the existence of 
$n$-times correlation probabilities $p^{t_1\cdots t_n}_{X_1\cdots X_n}$
satisfying positive definiteness and completeness, as well as the non-invasive 
measurability at any intermediate time \cite{LEGGETT}. Under these conditions,
for instance,
\begin{equation}
\label{DERIVTBI}
p^{\rm ac}_{+-}=p^{\rm abc}_{+--}+p^{\rm abc}_{++-}\leq
p^{\rm abc}_{+--}+p^{\rm abc}_{+-+}+p^{\rm abc}_{++-}+p^{\rm
abc}_{-+-}=p^{\rm ab}_{+-}+p^{\rm bc}_{+-}
\end{equation}
which involves measurements performed at the 
subsequent times $t_{\rm b}$ and $t_{\rm c}$ on a system prepared 
in state $|+ \rangle$ at the initial time $t_{\rm a}$. 
More in general three types of inequalities, analogous to 
(\ref{DERIVTBI}), can be derived:
\begin{eqnarray}
\Delta P_{\rm I}(t_{\rm ab},t_{\rm bc}) & \equiv & 
p^{\rm ac}_{X_{\rm a},X_{\rm c}} - p^{\rm ab}_{X_{\rm a},X_{\rm b}} - p^{\rm
bc}_{-X_{\rm b},X_{\rm c}}\leq 0\;;\nonumber\\
\Delta P_{\rm II}(t_{\rm ab},t_{\rm bc}) & \equiv & 
p^{\rm ab}_{X_{\rm a},X_{\rm b}} - p^{\rm ac}_{X_{\rm a},X_{\rm c}} - p^{\rm
bc}_{X_{\rm b},-X_{\rm c}}\leq 0\;;\label{TYPES}\\
\Delta P_{\rm III}(t_{\rm ab},t_{\rm bc}) & \equiv & 
p^{\rm bc}_{X_{\rm b},X_{\rm c}} - p^{\rm ab}_{X_{\rm a},X_{\rm b}} - p^{\rm
ac}_{-X_{\rm a},X_{\rm c}}\leq 0\;.\nonumber
\end{eqnarray}
While inequalities of type III are ruled out by the experimental requirement 
that the system is prepared in a definite state at time $t_{\rm a}$, the other
two
types are in principle well suitable for experimental test. In fact,
as originally observed in \cite{LEGGETT}, each one of them is violated for 
some subset of measurement times $t_{\rm ab}\equiv t_{\rm b}-t_{\rm a}$, 
$t_{\rm bc}\equiv t_{\rm c}-t_{\rm b}$.
On the other hand, the distinguishability criterion is fulfilled for another 
subset of $(t_{\rm ab},t_{\rm bc})$ depending upon the chosen threshold
effective uncertainty $\xi$. It 
is therefore useful to introduce an overlap integral expressing the 
average probability difference $\Delta P_\alpha$ (where 
$\alpha={\rm I},{\rm II},{\rm III}$) integrated over all the time 
intervals for which both temporal Bell inequalities are violated by quantum 
mechanical predictions and distinguishability is assured by 
$\Delta X_{\rm eff}\leq\xi$: 
\begin{equation}
{\cal O}_\alpha(\xi )\equiv\frac{1}{\tau^2}
\int_\cap \Delta P_\alpha\,{\rm d}t_{\rm ab}\,{\rm d}t_{\rm bc}
\label{OVERLAP}
\end{equation}
where the subscript $\cap$ means that the integration is restricted to the
values of $(t_{\rm ab},t_{\rm bc})$ for which $\Delta P_\alpha >0$ and 
$\Delta X_{\rm eff}\leq\xi $ for all measurements in $\Delta P_\alpha$.
The meaning of the overlap integral Eq.~(\ref{OVERLAP}) is the following:
given a threshold effective uncertainty $\xi$, it provides a measure of
whether the violation of a certain temporal Bell inequality is compatible with
the discrimination between the two values of the dichotomic observable,
according to the criterion defined by $\xi$. 
From Eq.~(\ref{OVERLAP}), it follows that ${\cal O}_\alpha(\xi)$ is 
continuous and monotonically increasing with $\xi$. 
Based on the considerations below
Eq.~(\ref{DSIGMAABBC}), we expect also ${\cal O}_\alpha(\xi\to 0)=0$, 
${\cal O}_\alpha(\xi\to 2|X|)=\tau^{-2}\int_0^\tau{\rm d}t_{\rm ab}
\int_0^\tau{\rm d}t_{\rm bc}\Delta P_\alpha>0$. Therefore the overlap integral 
will be nonzero only for $\xi$ large enough. We thus define
\begin{equation}
\xi_\alpha\equiv\min\{\xi:{\cal O}_\alpha(\xi')>0, \xi'>\xi\}=
\max\{\xi:{\cal O}_\alpha(\xi')<0, \xi'<\xi\}.
\end{equation}
This has to be compared with the $\xi_{1/2}$ defined above, in order to find out if 
violation of the inequalities Eq.~(\ref{TYPES}) and 
discrimination of the two values of $X$ can be obtained in the same experiment.
As a tool for our analysis we first discuss a toy model exhibiting all the 
relevant features intrinsic to the temporal Bell inequalities, moving later 
to the actual Hamiltonian of a rf-SQUID system as proposed in \cite{TESCHE}.
We do not take into account finite temperature or
imperfect efficiency effects, 
irrelevant to the arguments we plan to discuss since they 
can only diminish the chance to observe the predicted violations 
(see for instance the discussion in \cite{SANTOS}).

\section{Spin-1/2 particle case}
Let us consider a spin-1/2 
particle precessing in a uniform magnetic field. Its Hamiltonian is 
$\hat{H}=-\vec{\sigma} \cdot \vec{B}$ with $\vec{\sigma}=(\sigma_x, 
\sigma_y, \sigma_z)$ the Pauli matrices and $\vec{B}=(B,0,0)$. 
The temporal evolution of the state vector is ruled by the operator 
\begin{equation}
\hat{U}(t_{\rm b},t_{\rm a})=\exp\left[-\frac{i}{\hbar} 
B \sigma_x(t_{\rm b}-t_{\rm a})\right]\;.
\end{equation}
A measurement of the third component of the spin is represented by the
projector
\begin{equation}
\hat{\Pi}_{\Sigma}= {{1+2\Sigma\sigma_z} \over 2}\;,
\qquad \Sigma=\pm \frac 12\;.
\end{equation}
The correlation probabilities appearing in  
(\ref{DERIVTBI}) are then: 
\begin{equation}
\label{PROBTBI}
p^{\rm ab}_{+-}=\sin^2 \Omega t_{\rm ab}\;; \qquad
p^{\rm ac}_{+-}=\sin^2 \Omega (t_{\rm ab}+t_{\rm bc})\;; \qquad
p^{\rm bc}_{+-}=\cos^2 \Omega t_{\rm ab}\sin^2 \Omega t_{\rm bc}\;. 
\end{equation}
where the angular frequency $\Omega\equiv B/\hbar$ has been introduced, whence
the oscillation period $\tau_\sigma=2\pi/\Omega=2\pi\hbar/B$. The effective
uncertainties are then calculated according to (\ref{DSIGMAABBC}).
As already pointed out in \cite{KONST}, this discussion
applies to the proposals \cite{PAZ} and \cite{SANTOS}, 
dealing with the two-level dynamics of atomic systems. 

\section{SQUID case}
In the rf-SQUID case instead the monitored observable is the magnetic flux 
$\Phi$ threading the ring, a {\it continuous} observable.  
Its sign is a dichotomic variable which is directly measurable in a quantum 
coherence experimental setup \cite{DIAMBRINI}; the corresponding projector is 
\begin{equation}
\hat{\Pi}_\Phi =\Theta (\Phi\hat{\varphi})\;,\qquad\Phi=\pm 1\; .
\end{equation}
The effective Hamiltonian describing the system is \cite{BARONE}
\begin{equation}
\hat{H}= 
- {\hbar^2 \over 2C} {\drm^2 \over \drm \varphi^2} + 
{(\varphi-\Phi^{\rm ext})^2 \over 2L} - {I_{\rm c} \Phi_0 \over 2 \pi} \cos 
\left( 2\pi {\varphi \over \Phi_0 } \right) \;, 
\label{HAMSQ}
\end{equation}
where $C$ is the weak link capacitance (which plays the role of an effective 
mass), $L$ the ring inductance, $I_{\rm c}$ 
the critical current of the junction, $\Phi^{\rm ext}$ is the external 
magnetic flux threading the ring and the flux quantum $\Phi_0=\hbar/2e 
\approx 2.07\cdot 10^{-15}$ Wb. 
A bistable regime is obtained if the condition   
$1<2\pi LI_{\rm c}/\Phi_0<5\pi/2$ is fulfilled. In this case, by introducing the 
adimensional 
variable $\phi= (\varphi- \Phi^{\rm ext})/\Phi_0$ and by fixing the value 
of the external flux at $\Phi^{\rm ext}/\Phi_0=(n+1/2)$, $n\in$ {\bf Z} 
the potential can be approximated --~up to a constant~-- by the quartic
double-well form
\begin{equation}
V(\phi)= \frac{\pi^3}{3}I_{\rm c}\Phi_0(\phi-\phi_0)^2(\phi+\phi_0)^2
\;, \label{POT}
\end{equation}
where the two minima $\pm \phi_0$ are the solutions of the 
equation $\sin(2 \pi \phi)/\phi=\Phi_0/LI_{\rm c}$ and are separated by a barrier of 
height $\Delta V=(\pi^3/3)I_{\rm c}\Phi_0\phi_0^4$. Correspondingly, the 
ground state has two peaks localized 
around $\pm \phi_0$, each of width $\sigma_0$ determined by the approximate 
relationship $C\omega_0^2 \sigma_0^2/2 \simeq \hbar \omega_0/4$ 
where $\omega_0$ is the plasma frequency, {\it i.e.} the angular frequency of 
the small oscillations around $\pm \phi_0$. A flux wavepacket localized at 
$\pm \phi_0$ is a superposition of the ground and first excited state and 
has width $\sigma_0$. If no measurement is performed, such a state will 
undergo coherent tunneling oscillations between the two wells with period 
$\tau_\phi\equiv 2\pi\hbar/\Delta E_0$, where $\Delta E_0$ is the splitting 
between the lowest two energy eigenvalues. This last decreases exponentially 
with increasing barrier height, whereas the tunneling frequency should be at 
least of the order of magnitude of 1 MHz -- otherwise the coherent 
oscillations would be damped by the interaction with the environment. 
On the other hand, in
order to treat the system as a two-level one \cite{LEGGETT}, the probability 
for finding the system in the barrier region around $\phi=0$ should be 
negligible; in other words, it should be $\sigma_0^2\ll \phi_0^2$, which in 
turn could be achieved with a high enough barrier. It is possible 
to choose the parameters in (\ref{HAMSQ}) in order to satisfy both these
requirements: indeed, with $L=150 pH$, $C=0.15 pF$ and $I_{\rm c}=2.5\mu A$ (very
close to the values used in \cite{TESCHE}) it is 
$\tau_\phi^{-1}\approx 5.94\;\mbox{MHz}$ and $\sigma_0^2/\phi_0^2\approx 0.08$.

\section{Results and discussion}
\begin{figure}
\begin{center}
\epsfig{figure=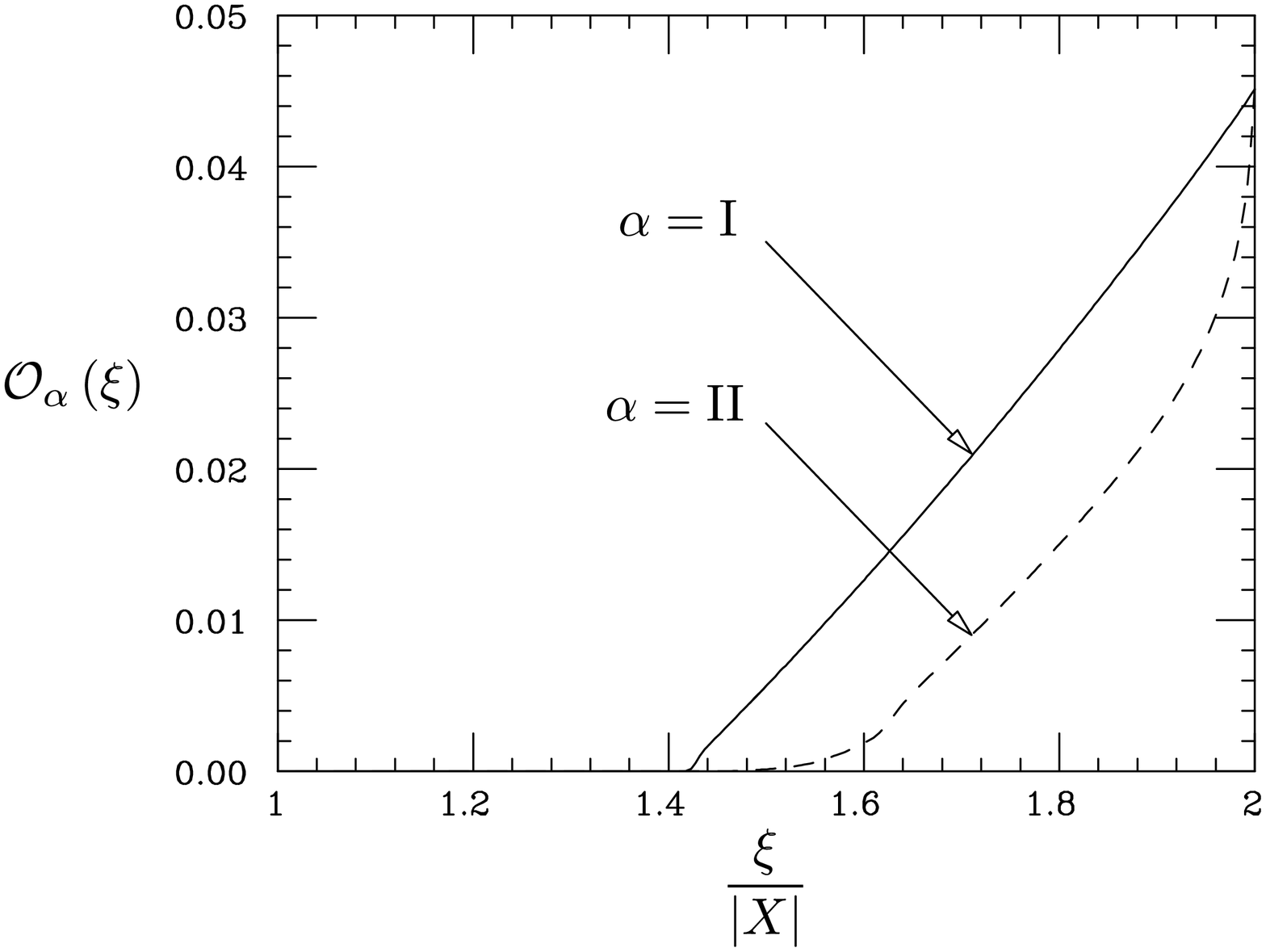,width=4in}
\end{center}
\caption{\label{fig1}
Overlap integral for inequalities of type I and II, as function of 
the effective uncertainty threshold $\xi$ (in units of $|X|$),
with any combinations of measurement results $X_a$, $X_b$ and $X_c$.
}
\end{figure}

In fig.~\ref{fig1} the overlap integral defined in Eq.~(\ref{OVERLAP}) is plotted
as function of $\xi $ for all the inequalities of the form 
(\ref{DERIVTBI}). 
The result depicted here is valid both 
for the spin toy-model and for
the actual SQUID Hamiltonian (with the parameters quoted
above), which lead to identical predictions because, as the tunneling 
oscillations dominate the dynamics, the two-level approximation is 
appropriate to describe the flux correlation probabilities in the rf-SQUID.
No dependence on the actual values of $X_{\rm a}$, $X_{\rm b}$ and $X_{\rm c}$ 
is seen, but only on the type of inequality. The type I turns out to be more 
favourable for the observation of violations.
Nevertheless, as it can be seen from the Figure, in both cases 
$\xi_\alpha>1.4|X|$, which is greater than any reasonable resolution threshold. 
This is the main result of this Letter 
showing that, even in the most favourable situation, it is impossible to 
observe violations to temporal Bell inequalities predicted by quantum mechanics 
maintaining at the same time the resolution high enough to distinguish 
the two values of the  dichotomic observable.

The correlation probabilities (\ref{PABPBC}) are obtained by 
means of the standard rules of quantum mechanics from the corresponding 
interfering amplitudes. 
The violation of the inequality (\ref{DERIVTBI}) however may be understood  by
noticing that it necessarily implies that at least one of the three time
correlation probabilities  $p_{ijk}^{\rm abc}$  becomes negative, namely becomes
a Wigner function (pseudoprobability). This means, according to Feynman
\cite{FEYNMAN}, that ``the assumed condition of preparation or verification are
experimentally unattainable'' as a consequence of the uncertainty
principle which forbids the existence of joint probabilities for
incompatible variables. For instance, in our case from (\ref{DERIVTBI}) follows that,
when $X_{\rm a}=-X_{\rm b}=-X_{\rm c}=|X|$ in (\ref{TYPES}), 
$\Delta P_{\rm I}=-p_{+-+}^{\rm abc}$: therefore in the region where the
temporal Bell inequality is violated, at least one of the joint
pseudoprobabilities at three different times is negative. This result holds in
general -- and is no surprise. One should remember in fact that Wigner's proof
\cite{WIGNER} of validity of spatial Bell inequalities is based on the assumption,
analogous to that made before Eq.~(\ref{DERIVTBI}), of non-negative joint
probabilities for the spin components along different directions. Their
violation should be ascribed, therefore, to the uncertainty principle which
prevents the existence of these non-negative joint probabilities. This
consequence of the uncertainty principle is independently confirmed by a
recent result \cite{CERF} according to which the violation of spatial Bell
inequalities is directly connected to the appearance of negative conditional
entropies, a feature which is classically forbidden.         

On the other hand, and this is the difference between spatial and temporal Bell
inequalities, it is the uncertainty principle itself which introduces in our
case a constraint on the ability to resolve the two states in a second
measurement at a later time on the same  particle. (Such a constraint does not
exist when the two measurements are performed at the same time on two
different particles located in different space points). This second constraint
arises from the fact that, as discussed in detail in \cite{ONCAL}, a quantum
measurement of a given observable induces a perturbation in the evolution of
its canonical conjugate, and this in turn produces an uncertainty on the
outcome of a measurement of the same variable at a later time. This is crucial
because, as we mentioned at the beginning of this paper, repeating
measurements on the same observable is precisely what discriminates
temporal Bell inequalities from the spatial version, where two different
systems are observed only once. These two constraints, both arising from
the uncertainty principle, are, as we have seen, conflicting, because
the region in which the violations arise and the region in which the
resolution is high enough to resolve between the two states do not
overlap.

These general considerations should affect other situations in which a 
macroscopic quantum system is repeatedly measured, for instance 
an atomic Bose-Einstein condensate in which the Josephson dynamics 
is studied. At the very core of quantum mechanics, the uncertainty 
principle has two competing consequences when applied to a single 
macroscopic system repeatedly monitored, the violation of classical 
probability laws for predictions on a dichotomic observable 
and the limitations on the resolution of the observable itself, 
and it seems impossible to experimentally unravel these two features 
without conflict.

%
%
\stars
We thank C.~Cosmelli and C.~Kuklewicz for useful discussions, and L.~Chiatti 
for comparison of codes simulating the behaviour of the rf-SQUID.
One of us (T.~C.) thanks the Institute of Nuclear Theory at the University of
Washington and the  Department of Physics at the University of Trento 
(in particular
M.~Traini and S.~Stringari) for their hospitality, the Department of Energy
and the ECT* and INFN in Trento 
for  partial support during the completion of this work, and
A.~J.~Leggett and A.~Garg for stimulating discussions. 
%
%

\end{document}

%% file: euromacr.tex
\def\etal{{\hbox{{\tenit\ et al.\/}\tenrm :\ }}}

\def\And{{\rm and\ }}

\def\drm{{\rm d}}

\def\stars{\bigskip\centerline{***}\medskip}

\newif\ifboo \boofalse

\def\Review#1{\boofalse{\it #1},}
\def\Name#1{{\sc #1},}
\def\Vol#1{\ifboo Vol. {\bf #1}\else{\bf #1}\fi}
\def\Year#1{\ifboo #1\else(#1)\fi}
\def\Book#1{\bootrue{\it #1},}
\def\Page#1{\ifboo {\rm p. #1}\else{\rm #1}\fi}